\documentclass[fleqn,a4paper]{iopart}
\pdfoutput=1

\expandafter\let\csname equation*\endcsname\relax 
\expandafter\let\csname endequation*\endcsname\relax 
\usepackage{amsmath,amssymb,bm,bbm,xcolor,mathdots,stmaryrd}
\usepackage[colorlinks=true, linkcolor=blue, citecolor=magenta, urlcolor=blue]{hyperref}

\makeatletter
\let\@fnsymbol\@arabic
\makeatother

\usepackage[T1]{fontenc}
\usepackage{lmodern}
\usepackage[utf8]{inputenc}

\usepackage{microtype}
\usepackage{orcidlink}

\usepackage[caption=false]{subfig}
\usepackage{graphicx}

\newcommand{\HH}{\mathcal{H}}
\newcommand{\PP}{\mathcal{P}}
\renewcommand{\AA}{\mathcal{A}}

\renewcommand{\SS}{\mathcal{S}}
\newcommand{\SU}{\mathrm{SU}}
\newcommand{\U}{\mathrm{U}}
\renewcommand{\d}{\mathrm{d}}
\newcommand{\bra}[1]{\langle #1|}
\newcommand{\ket}[1]{|#1\rangle}
\newcommand{\braket}[2]{\langle #1 | #2 \rangle}
\newcommand{\ketbra}[2]{|#1 \rangle\langle #2|}
\newcommand{\dt}{\mathrm{d}t}
\newcommand{\ds}{\mathrm{d}s}

\newcommand{\length}{\mathcal{L}}

\newcommand{\llangle}{\langle\hspace{-2pt}\langle}
\newcommand{\rrangle}{\rangle\hspace{-2pt}\rangle}

\newcommand{\gfs}{g_{\textsc{fs}}}
\renewcommand{\mod}{\operatorname{mod}}

\usepackage[margin=2.5cm]{geometry}

\begin{document}
\title[Tight lower bounds on the time it takes to generate a geometric phase]{\LARGE Tight lower bounds on the time it takes to generate a geometric phase}

\author{Niklas H{\"o}rnedal\,\orcidlink{0000-0002-2005-8694}\,}
\address{Department of Physics and Materials Science, University of Luxembourg, L-1511 Luxembourg, Luxembourg}
\ead{niklas.hornedal@uni.lu}
\author{Ole S{\"o}nnerborn\,\orcidlink{0000-0002-1726-4892}\,}
\address{Department of Mathematics and Computer Science, Karlstad University, 651 88 Karlstad, Sweden}
\address{Department of Physics, Stockholm University, 106 91 Stockholm, Sweden}
\ead{ole.sonnerborn@kau.se}


\begin{abstract}
Geometric phase is a concept of central importance in virtually every branch of physics. In this paper, we show that the evolution time of a cyclically evolving quantum system is restricted by the system's energy resources and the geometric phase acquired by the state. Specifically, we derive and examine three tight lower bounds on the time required to generate any prescribed Aharonov-Anandan geometric phase. The derivations are based on recent results on the geometric character of the Mandelstam-Tamm and Margolus-Levitin quantum speed limits.
\end{abstract}

\section{Introduction}
In 1984, Michael Berry reported a discovery proven to have a surprisingly wide range of applications. Berry \cite{Be1984} showed that if the Hamiltonian of a quantum mechanical system depends on external parameters that are varied cyclically in an adiabatic manner, then each nondegenerate eigenstate of the Hamiltonian acquires a phase only depending on the geometry of the parameter space. Nowadays, the Berry phase is a concept of central importance in virtually every branch of modern physics \cite{ShWi1989, BoMoKoNiZw2003}, including the recent fields of Topological states of matter \cite{BeHu2013, ChTeScRy2016, HaDu2017} and Quantum computation \cite{ZaRa1999, SjToAnHeJoSi2012, AlSj2022, ZhKyFiKwSjTo2021}.

A few years after the publication of reference \cite{Be1984}, Aharonov and Anandan \cite{AhAn1987} extended Berry's work by showing that a geometric phase can be associated with every cyclically evolving system, not only those evolving adiabatically. Although often referred to as a nonadiabatic phase, the Aharonov-Anandan geometric phase is also defined for adiabatically evolving systems and then agrees with the Berry phase.

The Aharonov-Anandan phase depends neither on the evolution time nor on the rate at which the system evolves. However, the path followed by a cyclically evolving state acquiring a nontrivial Aharonov-Anandan phase cannot be arbitrarily short. In this paper, we derive a lower bound for the Fubini-Study length of a closed curve of states in terms of its Aharonov-Anandan phase. Then, proceeding from the geometric interpretation of the Mandelstam-Tamm quantum speed limit  \cite{MaTa1945, AnAh1990}, we derive a tight lower bound on the time it takes to generate a specified Aharonov-Anandan phase. Interestingly, the Margolus-Levitin quantum speed limit \cite{MaLe1998} is also connected to the Aharonov-Anandan phase. Using a geometric description of the Margolus-Levitin quantum speed limit \cite{HoSo2023a}, we derive another tight lower bound on the time it takes to generate an Aharonov-Anandan phase.

Generally, a quantum speed limit is a fundamental estimate on the time required to transform a quantum system in a specified way \cite{Fr2016, DeCa2017}. As announced, the evolution time estimates derived here originate from geometric characterizations of the Mandelstam-Tamm and Margolus-Levitin quantum speed limits \cite{MaTa1945, MaLe1998, HoSo2023a, Fl1973, GiLlMa2003, HoAlSo2022, HoSo2023b}. We have organized the paper as follows. In Section \ref{sec: AA phase} we review the definition of the Aharonov-Anandan geometric phase, and in Section \ref{sec: MLQSL} we formulate and discuss some properties of a Margolus-Levitin type estimate for systems whose dynamics are driven by time-independent Hamiltonians. Estimates of Margolus-Levitin type do not extend straightforwardly to systems with time-dependent Hamiltonians \cite{HoSo2023b}, but the Mandelstam-Tamm quantum speed limit does \cite{AnAh1990}. In Section \ref{sec: MTQSL} we formulate an estimate of Mandelstam-Tamm type for time-dependent systems and relate the bound to the Margolus-Levitin type bound via the Bhatia-Davies inequality \cite{BhDa2000}. Section \ref{sec: Derivations} contains derivations of the evolution time estimates. The paper ends with a summary and an outlook.

\paragraph{Remark}
The problem addressed here is a kind of Brachistochrone problem \cite{CaHoKoOk2006, CaHoKoOk2007, CaHoKoOk2008, WaAlJaLlLuMo2015, AlHoAn2021}. However, unlike the case in most Brachistochrone problems, no constraints need to be placed on the spectrum of the Hamiltonian. This makes the bounds derived here universally valid.

\section{The Aharonov-Anandan phase}\label{sec: AA phase}
The Aharonov-Anandan geometric phase of a cyclic evolution of a pure state is the argument of the holonomy of the evolution in the gauge structure established by the Hopf bundle equipped with the Aharonov-Anandan connection \cite{AhAn1987, Si1993}. Here we briefly review this definition. For more detailed accounts consult reference \cite{BoMoKoNiZw2003} or reference \cite{ChJa2004}.

Consider a quantum system modeled on a finite-dimensional Hilbert space $\HH$. Let $\SS$ be the unit sphere in $\HH$, and $\PP$ be the projective Hilbert space of unit rank orthogonal projection operators of $\HH$; such operators represent pure states of the system.\textsuperscript{\footnotemark}\footnotetext{We will only consider quantum systems in pure states. The word ``state'' will, therefore, always refer to a pure quantum state.} The Hopf bundle is the $\U(1)$ principal bundle $\eta$ sending $\ket{\psi}$ in $\SS$ to $\ketbra{\psi}{\psi}$ in $\PP$. Two vectors $\ket{\psi}$ and $\ket{\phi}$ in $\SS$ belong to the same fiber of $\eta$, that is, get sent to the same state in $\PP$, if and only if $\ket{\phi}=e^{i\theta}\ket{\psi}$ for some relative phase $\theta$. We recommend reference \cite{KoNo1996} as a general reference on the theory of fiber bundles.

Suppose $\ket{\psi}$ sits in the fiber over $\rho$ in $\PP$. Let $\ket{\dot\psi}$ be a tangent vector at $\ket{\psi}$ and $\dot\rho$ be a tangent vector at $\rho$. We say that $\ket{\dot\psi}$ is a lift of $\dot\rho$ if the differential of $\eta$ maps $\ket{\dot\psi}$ onto $\dot\rho$. Similarly, we call a curve $\ket{\psi_t}$ in $\SS$ a lift of a curve $\rho_t$ in $\PP$ if $\eta$ maps $\ket{\psi_t}$ onto $\rho_t$. If so, each velocity vector $\ket{\dot\psi_t}$ is a lift of the corresponding velocity vector $\dot\rho_t$.

The Aharonov-Anandan connection is the real-valued $1$-form $\AA$ on $\SS$ whose action on a tangent vector $\ket{\dot\psi}$ at $\ket{\psi}$ is $\AA\ket{\dot\psi}=i\braket{\psi}{\dot\psi}$. We say that $\ket{\dot\psi}$ is horizontal if $\AA\ket{\dot\psi}=0$. The theory of principal bundles \cite{KoNo1996} tells us that if $\ket{\psi}$ sits in the fiber over $\rho$, then each tangent vector $\dot\rho$ at $\rho$ lifts to a unique horizontal vector $\ket{\dot\psi}$ at $\ket{\psi}$, and more generally that if $\rho_t$ is a curve extending from $\rho$, then $\rho_t$ lifts to a unique curve $\ket{\psi_t}$ that extends from $\ket{\psi}$ and is such that $\ket{\dot\psi_t}$ is horizontal for every $t$. The curve $\ket{\psi_t}$ is called the horizontal lift of $\rho_t$ starting at $\ket{\psi}$.

\begin{figure}[t]
	\centering
	\includegraphics[width=0.39\linewidth]{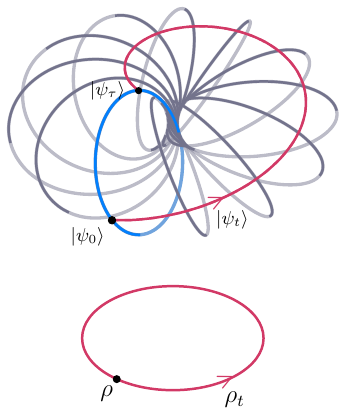}
    \caption{The figure shows the cyclic evolution of a qubit state $\rho_t$ and one of its horizontal lifts $\ket{\psi_t}$. The upper part of the figure displays the Hopf fibers over the states in the evolution curve, conformally mapped into Euclidean $3$-space by stereographic projection. The horizontal lift intersects each fiber perpendicularly, and the intersections form base points in the fibers allowing us to canonically identify the fiber over each $\rho_t$ with $\U(1)$ via $\ket{\phi}\mapsto \braket{\psi_t}{\phi}$. Since the initial and final vectors $\ket{\psi_0}$ and $\ket{\psi_\tau}$ belong to the same fiber, colored blue, we can identify $\braket{\psi_0}{\psi_\tau}$ with a phase factor $e^{i\gamma}$ via the identification of the blue fiber with $\U(1)$ associated with $\ket{\psi_0}$. This phase factor is the Aharonov-Anandan holonomy of $\rho_t$, and $\gamma$ is the Aharonov-Anandan geometric phase.}
	\label{fig: fas}
\end{figure}

Suppose that $\rho_t$, where $0\leq t\leq \tau$, is a closed curve at $\rho$. Let $\ket{\psi_t}$ 
be a horizontal lift of $\rho_t$. Since $\rho_t$ is closed, $\ket{\psi_t}$ starts and ends in the fiber over $\rho$; see Figure \ref{fig: fas}. The holonomy of $\rho_t$ is the relative phase factor $e^{i\gamma}=\braket{\psi_0}{\psi_\tau}$. 
The congruence class modulo $2\pi$ of the phase $\gamma$ is the Aharonov-Anandan geometric phase of $\rho_t$. Since the action by $\U(1)$ on $\SS$ takes horizontal curves to horizontal curves, the holonomy and the geometric phase do not depend on the choice of horizontal lift of $\rho_t$.

Although the geometric phase of $\rho_t$ is defined in terms of a horizontal lift, the geometric phase can be determined from an arbitrary lift: If $\ket{\psi_t}$ is any lift of $\rho_t$, then 
\begin{equation}\label{horizontal lift}
    \ket{\psi_t'}=
    \ket{\psi_t}\exp\Big(i\int_0^t\ds\,\AA\ket{\dot\psi_s}\Big)
\end{equation}
is a horizontal lift of $\rho_t$. The holonomy of $\rho_t$ is, thus, 
\begin{equation} 
    e^{i\gamma}
    =\braket{\psi_0}{\psi_\tau} \exp\Big(i\int_0^\tau\dt\,\AA\ket{\dot\psi_t}\Big).
\end{equation}
Consequently, 
\begin{equation}\label{geometric phase}
    \gamma = \arg\braket{\psi_0}{\psi_\tau}+i\int_0^\tau\dt\,\braket{\psi_t}{\dot\psi_t}\;\mod\,2\pi.
\end{equation}
Hereafter we will identify geometric phases with their representatives in the interval $[0,2\pi)$. Thus, when we write that $\rho_t$ has the geometric phase $\gamma$ we mean that $\gamma$ is the smallest nonnegative representative of the geometric phase of $\rho_t$.

\section{A Margolus-Levitin type estimate}\label{sec: MLQSL}
Consider a quantum system whose dynamics is driven by a time-independent Hamiltonian $H$. Suppose that the evolving state of the system forms a closed curve $\rho_t$, $0\leq t\leq \tau$, at $\rho$ with geometric phase $\gamma$. Write $\langle H \rangle$ for the expected energy, and let $\overline\epsilon$ be the largest occupied energy that is less than $\langle H \rangle$ and $\underline\epsilon$ be the smallest occupied energy that is greater than $\langle H \rangle$.\textsuperscript{\footnotemark}\footnotetext{An energy $\epsilon$ is occupied if there is a nonzero probability that $\epsilon$ is obtained at an energy measurement. Equivalently, an energy $\epsilon$ is occupied by $\rho$ if $\bra{\epsilon}\rho\ket{\epsilon}>0$ for an energy eigenstate $\ket{\epsilon}$ with eigenvalue $\epsilon$.} Then\textsuperscript{\footnotemark}\footnotetext{All quantities are expressed in units such that $\hbar=1$.}\textsuperscript{,}\textsuperscript{\footnotemark}\footnotetext{If $\overline\epsilon$ or $\underline\epsilon$ does not exist, the state is stationary, and we define the right-hand side of \eqref{MLQSL} to be $0$.}
\begin{equation}\label{MLQSL}
    \tau\geq
    \max\bigg\{ \frac{\gamma}{\langle H-\overline\epsilon\,\rangle}, \frac{2\pi-\gamma}{\langle\,\underline\epsilon-H \rangle}\bigg\}.
\end{equation}
The quotients on the right resemble the Margolus-Levitin quantum speed limit \cite{MaLe1998} and its dual \cite{NeAlSa2022, HoSo2023a}. Therefore, we call the bound on the right a bound of Margolus-Levitin type. The crucial observation leading to \eqref{MLQSL} is that if $\epsilon$ is an occupied energy, then
\begin{equation}\label{crucial observation}
    \gamma = \tau\langle H-\epsilon\rangle\;\mod\,2\pi.
\end{equation}
We derive the Margolus-Levitin type bound \eqref{MLQSL} in Section \ref{Derivation of MLQSL}.

\subsection{Tightness of the Margolus-Levitin type estimate}\label{ML tightness}
Here we show that for every value of $\gamma$, the estimate \eqref{MLQSL} can be saturated by a qubit. If $\gamma=0$, the estimate is saturated by a stationary state. We thus assume that $\gamma>0$.

Consider a system modeled on a Hilbert space spanned by orthonormal vectors $\ket{0}$ and $\ket{1}$. Define the Pauli operators as
\begin{align}
    &\sigma_x=\ketbra{0}{1}+\ketbra{1}{0}, \\
    &\sigma_y=i(\ketbra{0}{1}-\ketbra{1}{0}), \\
    &\sigma_z=\ketbra{1}{1}-\ketbra{0}{0}.
\end{align}
We can identify the projective Hilbert space with the unit sphere in Euclidean $3$-space by identifying each qubit state $\rho$ with a unit vector $\mathbf{r}$ defined as
\begin{equation}
    \mathbf{r}=\big(\!\tr(\rho\sigma_x),\tr(\rho\sigma_y),\tr(\rho\sigma_z)\big).
\end{equation}
The vector $\mathbf{r}$ is called the Bloch vector of $\rho$, and the unit sphere is called the Bloch sphere. One can also assign a vector $\mathbf{\Omega}$ to each qubit Hamiltonian $H$ called the Rabi vector of the Hamiltonian:
\begin{equation}
    \mathbf{\Omega}=\big(\!\tr(H\sigma_x),\tr(H\sigma_y),\tr(H\sigma_z)\big).
\end{equation}

\begin{figure}[t]
	\centering
	\includegraphics[width=0.39\linewidth]{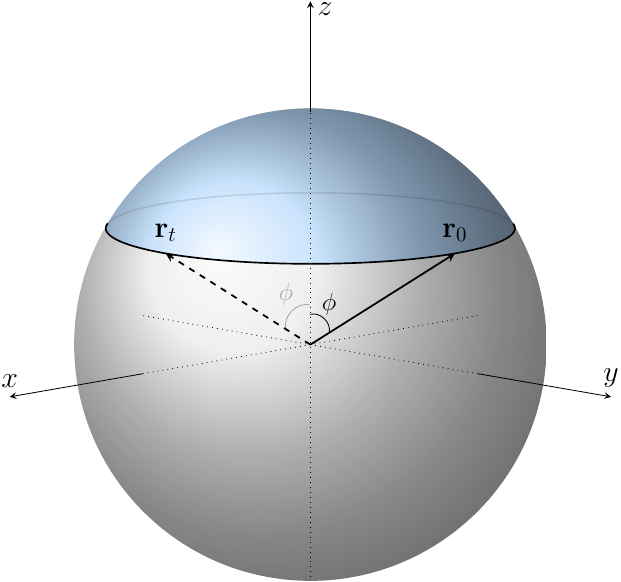}
	\caption{The evolving Bloch vector $\mathbf{r}_t$ rotates about the $z$-axis with a preserved polar angle and an azimuthal angular speed $\Omega$. Thus, it returns to its original position at $\tau=2\pi/\Omega$. The geometric phase equals $2\pi$ minus half the solid angle subtended by $\mathbf{r}_t$. The solid angle is the area of the blue spherical cap.}
	\label{fig: qubit}
\end{figure}

Assume that the Rabi vector of the system Hamiltonian $H$ points along the positive $z$-axis and has the length $\Omega$. In other words, assume that $\ket{0}$ and $\ket{1}$ represent the ground and excited states of $H$ and that $\Omega$ is the difference between the excited and ground state energies. Furthermore, assume that the system is prepared in a state $\rho$ whose Bloch vector $\mathbf{r}$ makes a polar angle $\phi>0$ with the positive $z$-axis; see Figure \ref{fig: qubit}.
As time passes, the Bloch vector $\mathbf{r}_t$ of the evolving state $\rho_t$ rotates around the $z$-axis with a preserved polar angle and azimuthal angular speed $\Omega$. The Bloch vector thus returns to its original position, and $\rho_t$ returns to $\rho$, at time $\tau=2\pi/\Omega$. One can show \cite{LaHuCh1990} that the geometric phase $\gamma$ of $\rho_t$ is equal to $2\pi$ minus half the solid angle subtended by $\mathbf{r}_t$,
\begin{equation}\label{tata}
    \gamma = \pi(1+\cos\phi).
\end{equation}
By adjusting $\phi$, the phase $\gamma$ can be given any predetermined value.

The eigenvalues of $H$ are $(\tr H\pm \Omega)/2$, both being occupied at all times, and the expected energy is $(\tr H+\Omega\cos\phi)/2$. Consequently,
\begin{align}
    &\langle H - \overline\epsilon\,\rangle 
    = \frac{\Omega}{2}(1+\cos\phi), \\
    &\langle\, \underline\epsilon - H\rangle 
    = \frac{\Omega}{2}(1-\cos\phi).
\end{align}
It follows that
\begin{equation}
    \frac{\gamma}{\langle H - \overline\epsilon\,\rangle}
    =\frac{2\pi-\gamma}{\langle\, \underline\epsilon - H\rangle}
    =\frac{2\pi}{\Omega}
    =\tau.
\end{equation}
This shows that the Margolus-Levitin type bound \eqref{MLQSL} is tight.

\subsection{On the period of local cyclic evolutions of pairs of maximally entangled qubits}
Entanglement is an important resource for quantum information processing. For bipartite systems, the degree of entanglement of a pure state is quantified by the joint entropy of the marginal states \cite{BeBePoSc1996}. The degree of entanglement cannot be changed by applying local invertible operations, e.g., local unitary rotations. Here we consider periodic evolutions of maximally entangled pairs of qubits generated by local traceless Hamiltonians with arbitrary but predetermined spectra. We show that the periods of such evolutions are quantized. Furthermore, we show that in a sense it takes twice as long to generate a homotopically trivial such evolution as it does to generate a homotopically nontrivial one.

All maximally entangled pairs of qubits can be written in the form 
\begin{equation}
    \rho=\ketbra{a,b}{a,b},\quad \ket{a,b}=\frac{1}{\sqrt{2}}\big(a\ket{00}+b\ket{01}-b^*\ket{10}+a^*\ket{11}\big),
\end{equation} 
where $a$ and $b$ are arbitrary complex numbers satisfying $|a|^2+|b|^2=1$ \cite{MiMo2003,Mi2006}. The map sending $\ketbra{a,b}{a,b}$ to $[\Re a\!:\!\Im a\!:\!\Re b\!:\!\Im b]$ topologically identifies the space of all such states with the $3$-dimensional real projective space.\textsuperscript{\footnotemark}\footnotetext{The $3$-dimensional real projective space is the $3$-dimensional unit sphere with antipodal points identified. We denote the equivalence class of a point $(x_1,x_2,x_3,x_4)$ on the $3$-sphere by $[x_1\!:\!x_2\!:\!x_3\!:\!x_4]$.} Since the fundamental group of the latter space consists of two elements, so does the fundamental group of the former. This means that there are two classes of closed curves of maximally entangled pairs of qubits, one whose members can be continuously contracted to a stationary curve (we call these homotopically trivial) and one whose members cannot be continuously contracted to a stationary curve (these are the homotopically nontrivial ones).

Representatives of the two classes can be generated with local $\SU(2)$ rotations,
\begin{equation}
    \ketbra{a_t,b_t}{a_t,b_t}=e^{-itH_1}\otimes e^{-itH_2}\ketbra{a,b}{a,b} e^{itH_1}\otimes e^{itH_2}.
\end{equation}
Here, $H_1$ and $H_2$ are traceless Hamiltonians that act on the first and second subsystem, respectively \cite{MiMo2003}. For a cyclic evolution generated by such rotations, the geometric phase $\gamma$ is $0$ or $\pi$. This follows from the observations that the second term in equation \eqref{geometric phase} vanishes due to $H_1$ and $H_2$ being traceless and that 
$\ketbra{a_\tau,b_\tau}{a_\tau,b_\tau}=\ketbra{a,b}{a,b}$ implies $\braket{a,b}{a_\tau,b_\tau}=\pm 1$. If $H_1$ and $H_2$ have spectra $\{-\epsilon_1,\epsilon_1\}$ and $\{-\epsilon_2,\epsilon_2\}$, with $\epsilon_1$ and $\epsilon_2$ being positive, 
equation \eqref{crucial observation} tells us that 
\begin{equation}
    \tau(\epsilon_1+\epsilon_2)=0\;\mod \pi.
\end{equation}
Hence, for predetermined $\epsilon_1$ and $\epsilon_2$, the period is quantized.

It is known that if $\gamma=0$, the evolution curve is homotopically trivial, and if $\gamma=\pi$, the evolution curve is homotopically nontrivial \cite{MiMo2003}. If the Hamiltonians $H_1$ and $H_2$ generate a homotopically nontrivial closed evolution in the time $\tau$, they generate a homotopically trivial closed evolution in the time $2\tau$. (This follows from the group structure of the fundamental group.) We ask whether, with preserved local energy spectra, it is possible to locally generate a nonstationary topologically trivial closed evolution of a maximally entangled pair of qubits in less than twice the time it takes to generate a topologically nontrivial evolution of such a pair. The Margolus-Levitin type bound says that this is not possible. Because according to this bound, it takes at least the time 
\begin{equation}
    \max\bigg\{\frac{\pi}{\langle H_1+H_2+\epsilon_1+\epsilon_2\rangle},\frac{2\pi-\pi}{\langle\epsilon_1+\epsilon_2-H_1-H_2\rangle}\bigg\}=\frac{\pi}{\epsilon_1+\epsilon_2}
\end{equation}
to generate a homotopically nontrivial closed evolution, and at least the time 
\begin{equation}
    \max\bigg\{0,\frac{2\pi}{\langle\epsilon_1+\epsilon_2-H_1-H_2\rangle}\bigg\}=\frac{2\pi}{\epsilon_1+\epsilon_2}
\end{equation}
to generate a homotopically trivial nonstationary closed evolution. The estimates are tight and can be met using, e.g., $H_1=\epsilon(\ketbra{1}{1}-\ketbra{0}{0})$, where $\epsilon>0$, and $H_2=0$.

\subsection{An example that contrasts with the case of the Margolus-Levitin quantum speed limit}
A quantum mechanical system with a time-independent Hamiltonian that returns to its original state at some point has periodic dynamics. Section \ref{ML tightness} shows that every qubit system with a time-independent Hamiltonian is periodic, with a period given by the equal quotients on the right-hand side of \eqref{MLQSL}. In this section, we show that the corresponding statement does not hold in higher dimensions. More precisely, we show that \eqref{MLQSL} may but need not be the period of a qutrit system with all three energy levels occupied. Furthermore, we show that if the right-hand side of \eqref{MLQSL} is the period of the qutrit, then either of the quotients can be greater than the other. That the Margolus-Levitin type bound \eqref{MLQSL} can be saturated by a system with an effective dimension greater than two contrasts with the case of the Margolus-Levitin quantum speed limit 
that can only be saturated with effective qubits \cite{HoSo2023a}.

Consider a qutrit system with a Hamiltonian $H$ having eigenvalues $\epsilon_0<\epsilon_1<\epsilon_2$. Assume that the three eigenvalues are occupied and the evolution of the state is periodic with period $\tau$. Let $\gamma$ be the geometric phase of the path followed by the state during one period. According to \eqref{crucial observation}, there exist unique integers $n_0>n_1>n_2$ such that $\gamma=\tau\langle H-\epsilon_j\rangle-2\pi n_j$.

If the state is such that $\langle H\rangle<\epsilon_1$, then
\begin{align}
    &\frac{\gamma}{\langle H-\overline\epsilon\,\rangle}
    =\frac{\tau\langle H-\epsilon_0\rangle-2\pi n_0}{\langle H-\epsilon_0\rangle}
    =\tau-\frac{2\pi n_0}{\langle H-\epsilon_0\rangle},\label{nitton} \\
    &\frac{2\pi-\gamma}{\langle\,\underline\epsilon - H \rangle}
    =\frac{\tau\langle\epsilon_1-H\rangle+2\pi(n_1+1)}{\langle\epsilon_1- H \rangle}
    =\tau+\frac{2\pi(n_1+1)}{\langle\epsilon_1- H \rangle},\label{tjugo}
\end{align}
and if $\langle H\rangle>\epsilon_1$, then
\begin{align}
    &\hspace{-1pt}\frac{\gamma}{\langle H-\overline\epsilon\,\rangle}
    = \frac{\tau\langle H-\epsilon_1\rangle-2\pi n_1}{\langle H-\epsilon_1\rangle}
    =\tau-\frac{2\pi n_1}{\langle H-\epsilon_1\rangle},\label{tjuett} \\
    &\hspace{-1pt}\frac{2\pi-\gamma}{\langle\,\underline\epsilon - H \rangle}
    =\frac{\tau\langle \epsilon_2-H\rangle+2\pi(n_2+1)}{\langle\epsilon_2- H \rangle}
    =\tau+\frac{2\pi(n_2+1)}{\langle\epsilon_2- H \rangle}.\label{tjutva}
\end{align}
In both cases, with suitable choices of initial states and  eigenvalues, we can arrange it so that none, either, or both of the quotients on the right in the Margolus-Levitin type estimate \eqref{MLQSL} assume the value $\tau$. If, on the other hand, $\langle H\rangle=\epsilon_1$, then $\gamma=0$ and $n_1=0$. In this case, the first quotient in \eqref{MLQSL} vanishes and the second quotient is
\begin{equation}
    \frac{2\pi-\gamma}{\langle\,\underline\epsilon - H \rangle}
    =\frac{2\pi}{\epsilon_2-\epsilon_1}
    =\frac{\tau}{-n_2}.
\end{equation}
We can arrange it so that $n_2=-1$ as well as $n_2<-1$.

Let us demonstrate our claims in the case when $\langle H\rangle<\epsilon_1$. Assume for simplicity that $\epsilon_0=0$, $\epsilon_1=\epsilon$, and $\epsilon_2=q\epsilon$ with $q$ rational. The periodicity requirement implies that $e^{-i\tau\epsilon_j}=e^{i\theta}$
for some phase $\theta$. Thus, the period is $\tau=2\pi m/\epsilon$ where $m$ is the smallest positive integer such that $qm$ is an integer. Also, $\tau\langle H\rangle-2\pi n_0=\tau\langle H-\epsilon\rangle-2\pi n_1$ and hence $n_0=m+n_1$. Then,  
\begin{align}
    &\frac{\gamma}{\langle H-\overline\epsilon\,\rangle}
    =\tau-\frac{2\pi n_0}{\langle H\rangle}, \\
    &\frac{2\pi-\gamma}{\langle\,\underline\epsilon - H \rangle}
    =\tau+\frac{2\pi(n_0-m+1)}{\langle\epsilon - H \rangle}.
\end{align}

First, suppose that $q$ is an integer. Then $m=1$ and $\tau=2\pi/\epsilon$. The assumption $\langle H\rangle<\epsilon$ implies that $\tau\langle H\rangle<2\pi$, and hence that $n_0=0$. In this case, both quotients are equal to the period,
\begin{equation}
    \frac{\gamma}{\langle H-\overline\epsilon\,\rangle}
    =\frac{2\pi-\gamma}{\langle\,\underline\epsilon - H \rangle}
    =\tau.
\end{equation}
Next, suppose $q$ is not an integer. Then $m\geq 2$. If we choose the initial state such that the occupation of $\epsilon_0$ is large enough for $\tau\langle H\rangle=2\pi m \langle H\rangle/\epsilon < 2\pi$, then $n_0=0$ and
\begin{equation}
    \frac{\gamma}{\langle H-\overline\epsilon\,\rangle}
    =\tau, \qquad
    \frac{2\pi-\gamma}{\langle\,\underline\epsilon - H \rangle}
    =\tau+\frac{2\pi(1-m)}{\langle\epsilon - H \rangle}\ne \tau.
\end{equation}
If we instead choose the state such $\langle H\rangle\approx \epsilon$, then $n_0=m-1$
and
\begin{equation}
    \frac{\gamma}{\langle H-\overline\epsilon\,\rangle}
    =\tau-\frac{2\pi (m-1)}{\langle H\rangle}\ne\tau, \qquad
    \frac{2\pi-\gamma}{\langle\,\underline\epsilon - H \rangle}
    =\tau.
\end{equation}
Finally, suppose $q$ is such that $m=3$ and choose the state such that $\langle H\rangle=\epsilon/2$. Then $\tau\langle H\rangle=3\pi$ and $n_0=1$. In this case, none of the quotients is equal to the period:
\begin{equation}
    \frac{\gamma}{\langle H-\overline\epsilon\,\rangle}
    =\tau-\frac{2\pi}{\langle H\rangle}\ne\tau, \qquad
    \frac{2\pi-\gamma}{\langle\,\underline\epsilon - H \rangle}
    =\tau-\frac{2\pi}{\langle\epsilon - H \rangle}\ne\tau.
\end{equation}

\paragraph{Remark} The qutrit example above shows that it is more challenging to characterize the systems that saturate the Margolus-Levitin type estimate \eqref{MLQSL} than those that saturate the Margolus-Levitin quantum speed limit, which can only be saturated with effective qubits \cite{HoSo2023a}. Furthermore, the example shows that either of the two quotients on the right-hand side of \eqref{MLQSL} can be larger than the other. This means that none of the quotients are redundant. This is analogous to the case of the Margolus-Levitin quantum speed limit and its dual \cite{HoSo2023a,NeAlSa2022}. Finally, the example provides support for the intuitive idea that which of the two quotients in \eqref{MLQSL} is larger depends on whether the expected energy is close to $\overline\epsilon$ or $\underline\epsilon$.

\subsection{Nonextensibility of the Margolus-Levitin type estimate to time-dependent systems}\label{sec: nonextensibility}
In reference \cite{HoSo2023b} it was demonstrated that the Margolus-Levitin quantum speed limit, and its generalization to an arbitrary fidelity \cite{HoSo2023a,GiLlMa2003}, does not straightforwardly generalize to an evolution time bound for systems with a time-dependent Hamiltonian. Here we similarly show that \eqref{MLQSL} does not generalize to a time bound for time-dependent systems. More specifically, we show that it is in general not possible to replace the denominators in \eqref{MLQSL} with the corresponding time averages by showing that there are systems that violate such an estimate.

Consider a qubit system prepared in the $\sigma_x$-eigenstate
\begin{equation}
    \rho=\frac{1}{2}\big(\ketbra{0}{0}+\ketbra{0}{1}+\ketbra{1}{0}+\ketbra{1}{1}\big).
\end{equation}
Fix an $E>0$, let $\mu(\chi)=E/(1-\cos\chi)$ for some $0<\chi<\pi/2$, and assume that the dynamics of the qubit is driven by the time-dependent Hamiltonian $H_t=e^{-iAt}He^{iAt}$ where
\begin{align}
    & A= \mu(\chi)\sin\chi\, \sigma_z,\\
    & H=\mu(\chi)(\sin\chi\, \sigma_z- \cos\chi\,\sigma_x).
\end{align}
Write $\rho_t$ for the state at time $t$ generated from $\rho$ by $H_t$. 

The time propagator associated with $H_t$ is $e^{-iAt}e^{-i(H-A)t}$; see reference \cite{HoSo2023b}. Since $\rho$ commutes with $H-A$, we have $\rho_t=e^{-iAt}\rho e^{iAt}$. Furthermore, since $A$ is proportional to $\sigma_z$, the Bloch vector of $\rho_t$, which initially points in the direction of the $x$-axis, moves along the equator on the Bloch sphere.
The speed of the Bloch vector is $2E\cot(\chi/2)$, and it thus returns to its original position, and $\rho_t$ returns to $\rho$, at 
\begin{equation}
    \tau=\frac{\pi}{E\cot(\chi/2)}.
\end{equation}

The spectrum of $H_t$ is time-independent and consists of the eigenvalues $-\mu(\chi)$ and $\mu(\chi)$, both being occupied at all times. Moreover, the expected energy has the constant value $-\mu(\chi)\cos\chi$. As a consequence, the quantities $\langle H_t - \overline\epsilon_t\, \rangle$ and $\langle\, \underline\epsilon_t - H_t\rangle$ are also constant with the values
\begin{align}
    &\langle H_t - \overline\epsilon_t\, \rangle = -\mu(\chi)\cos\chi +\mu(\chi)=E, \\
    &\langle\, \underline\epsilon_t - H_t\rangle =\mu(\chi)+\mu(\chi)\cos\chi= E\cot^2(\chi/2).
\end{align}
According to \eqref{tata}, the geometric phase of $\rho_t$ is $\gamma=\pi$, and since $\cot(\chi/2)>1$,
\begin{equation}\label{MLQSL2}
    \max\bigg\{ \frac{\gamma}{\llangle\langle H_t-\overline{\epsilon}_t\rangle\rrangle}, \frac{2\pi-\gamma}{\llangle\langle\, \underline{\epsilon}_t - H_t \rangle\rrangle}\bigg\}
    =
    \max\bigg\{ \frac{\pi}{E}, \frac{\pi}{E\cot^2(\chi/2)}\bigg\}=
    \frac{\pi}{E}.
\end{equation}
The double angular brackets denote time averages of the instantaneous expected values over $[0,\tau]$. As $\pi/E$ is greater than $\tau$, the expression on the left-hand side of equation \eqref{MLQSL2} is, in general, not an evolution time bound. This shows that the Margolus-Levitin bound \eqref{MLQSL} does not extend straightforwardly to systems with time-dependent Hamiltonians.

\section{A Mandelstam-Tamm type estimate}\label{sec: MTQSL}
Another time bound that is also valid for time-dependent systems can be derived from the fact that the evolution time multiplied by the time average of the energy uncertainty is equal to the Fubini-Study length of the evolution curve.

The Fubini-Study metric on the projective Hilbert space $\PP$ is induced from half of the Hilbert-Schmidt metric on the space of Hermitian operators on $\HH$,
\begin{equation}
    \gfs(\dot\rho_1,\dot\rho_2)=\frac{1}{2}\tr(\dot\rho_1\dot\rho_2).
\end{equation}
If $\rho_t$, where $0\leq t\leq \tau$, is generated by a Hamiltonian $H_t$, that is, if $\dot\rho_t=-i[H_t,\rho_t]$, the instantaneous Fubini-Study speed of $\rho_t$ squared equals the instantaneous energy variance:
\begin{equation}\label{fubini-study-uncertainty}
    \gfs(\dot\rho_t,\dot\rho_t)
    =\frac{1}{2}\tr\big((-i[H_t,\rho_t])^2\big)=\Delta^2H_t.
\end{equation}
The Fubini-Study length of $\rho_t$ thus equals the evolution time multiplied with 
the time average of the energy uncertainty over the evolution time interval $[0,\tau]$:
\begin{equation}
    \length(\rho_t)
    =\int_0^\tau\dt\sqrt{\gfs(\dot\rho_t,\dot\rho_t)}    
    =\int_0^\tau\dt\, \Delta H_t     
    = \tau\llangle \Delta H_t\rrangle.
\end{equation}
In Section \ref{Derivation of MTQSL} we show that if $\rho_t$ is a closed curve with geometric phase $\gamma$, then
\begin{equation}\label{tjunio}
    \length(\rho_t)
    \geq \sqrt{\gamma(2\pi-\gamma)}.
\end{equation}
As a consequence,
\begin{equation}\label{MTQSL}
    \tau \geq \frac{\sqrt{\gamma(2\pi-\gamma)}}{\llangle \Delta H_t\rrangle}.
\end{equation}
We call the estimate in \eqref{MTQSL} a time estimate of Mandelstam-Tamm type due to its similarity with the Mandelstam-Tamm quantum speed limit \cite{MaTa1945, Fl1973}.

\paragraph{Remark}
The Mandelstam-Tamm type estimate is also valid for systems with a time-independent Hamiltonian. Thus, for systems with time-independent Hamiltonians we can expand \eqref{MLQSL} to
\begin{equation}
    \tau\geq\max\bigg\{\frac{\gamma}{\langle H - \overline\epsilon\,\rangle},\frac{2\pi-\gamma}{\langle\, \underline\epsilon - H\rangle},\frac{\sqrt{\gamma(2\pi-\gamma)}}{\Delta H}\bigg\}.
\end{equation}

\subsection{Tightness of the Mandelstam-Tamm type estimate}
The map sending a qubit state onto its Bloch vector is a diffeomorphism from the projective Hilbert space onto the Bloch sphere. However, the map is not an isometry. Rather, the Fubini-Study metric pushed forward to the Bloch sphere equals a quarter of the spherical metric. Consequently, the Fubini-Study length of a curve on the Bloch sphere is half its spherical length.

Consider the system studied in Section \ref{ML tightness}. We shall show that it also satisfies the inequality in \eqref{MTQSL}. Since the radius of the curve traced by the Bloch vector is $\sin\phi$, the Fubini-Study length of the curve is $\pi\sin\phi$. Also, according to \eqref{tata}, the geometric phase is $\gamma = \pi(1+\cos\phi)$. It follows that
\begin{equation}
    \length(\rho_t)
    =\pi\sqrt{1-\cos^2\phi}
    =\sqrt{\gamma(2\pi-\gamma)}.
\end{equation}
Since the Fubini-Study length is the smallest possible for a closed evolution with geometric phase $\gamma$, the estimate \eqref{MTQSL} is saturated. Explicitly, $\Delta H=\Omega\sin\phi/2$ and hence
\begin{equation}
    \frac{\sqrt{\gamma(2\pi-\gamma)}}{\Delta H}
    =\frac{2\pi\sin\phi}{\Omega\sin\phi}
    =\frac{2\pi}{\Omega}
    =\tau.
\end{equation}

\subsection{A qubit example}
In Section \ref{sec: nonextensibility}, we constructed a qubit system that violates a hypothetical extension of the Margolus-Levitin type estimate to time-dependent systems. Let us show that this system obeys and saturate the Mandelstam-Tamm type estimate. 

The expected values for $H_t$ and $H_t^2$ are time-independent with the values $\langle H_t\rangle=-\mu(\chi)\cos\chi$ and $\langle H_t^2\rangle=\mu(\chi)^2$, respectively. Thus, the energy variance is preserved with the value
\begin{equation}
    \Delta H_t=\sqrt{\langle H_t^2\rangle-\langle H_t\rangle^2}=\mu(\chi)\sin\chi.
\end{equation}
The state returns for the first time to its original position at $\tau=\pi/E\cot(\chi/2)$, and the geometric phase of the evolution curve is $\gamma=\pi$. Consequently,
\begin{equation}
    \frac{\sqrt{\gamma(2\pi-\gamma)}}{\llangle\Delta H_t\rrangle}
    = \frac{\pi}{\mu(\chi)\sin\chi}
    = \frac{\pi(1-\cos\chi)}{E\sin\chi}
    = \frac{\pi}{E\cot(\chi/2)}
    =\tau.
\end{equation}

\subsection{A time-bound of Bhatia-Davies type}
In time-independent systems where the state occupies only two energy levels, the Mandelstam-Tamm type bound \eqref{MTQSL} is the geometric mean of the quotients on the right-hand side of the Margolus-Levitin type bound \eqref{MLQSL}. This follows from the Bhatia-Davies inequality \cite{BhDa2000}, stating that the variance of an observable $A$ is bounded from above according to
\begin{equation}
    \Delta^2 A\leq 
    \langle A-a_{\mathrm{min}}\rangle \langle a_{\mathrm{max}}-A\rangle,
\end{equation}
where $a_{\mathrm{min}}$ is the smallest and $a_{\mathrm{max}}$ is the largest occupied eigenvalue of $A$. The Bhatia-Davies inequality is an equality if and only if at most two of the eigenvalues of $A$ are occupied.

Consider a time-dependent system with Hamiltonian $H_t$ that returns to its initial state at time $\tau$ and then has acquired the geometric phase $\gamma$. By the Bhatia-Davies inequality,
\begin{equation}\label{timeBD}
    \llangle \Delta H_t\rrangle 
    \leq \Big{\langle}\hspace{-4pt}\Big{\langle} \sqrt{\langle H_t-\epsilon_{\mathrm{min};t}\rangle \langle \epsilon_{\mathrm{max};t}-H_t\rangle}\,\Big{\rangle}\hspace{-4pt}\Big{\rangle},
\end{equation}
where $\epsilon_{\mathrm{min};t}$ and $\epsilon_{\mathrm{max};t}$ are the smallest and largest occupied instantaneous energy levels. Equations \eqref{MTQSL} and \eqref{timeBD} imply
\begin{equation}\label{BDQSL}
    \tau\geq 
    \frac{\sqrt{\gamma(2\pi-\gamma)}}{\llangle\sqrt{\langle H_t-\epsilon_{\mathrm{min};t}\rangle \langle \epsilon_{\mathrm{max};t}-H_t\rangle}\,\rrangle}.
\end{equation}
We call this a time estimate of Bhatia-Davies type. Since, in general, the inequality \eqref{timeBD} is strict, the bound in \eqref{BDQSL} is weaker than the Mandelstam-Tamm type bound \eqref{MTQSL}. For qubits, the estimates \eqref{MTQSL} and \eqref{BDQSL} are saturated simultaneously.

\section{Derivations of the Margolus-Levitin and Mandelstam-Tamm type bounds}\label{sec: Derivations}
Here we derive the Margolus-Levitin and Mandelstam-Tamm type bounds \eqref{MLQSL} and \eqref{MTQSL}. In the derivations we use concepts, notation, and theory introduced in Section \ref{sec: AA phase}.

\subsection{Derivation of the Margolus-Levitin type bound}\label{Derivation of MLQSL}
Consider a quantum system with Hamiltonian $H$ and evolving state
$\rho_t$, where $0\leq t\leq \tau$, forming a closed curve at $\rho$ with geometric phase $\gamma$. To prove \eqref{MLQSL}, let $\epsilon$ be any occupied eigenvalue of $H$ and $\ket{\psi}$ be any vector in the fiber over $\rho$. Then, as we shall see, the solution to $\ket{\dot\psi_t}=-i(H-\epsilon)\ket{\psi_t}$ with initial condition $\ket{\psi_0}=\ket{\psi}$ is a closed lift $\ket{\psi_t}$ of $\rho_t$. According to \eqref{geometric phase},
\begin{equation}\label{geometric phase 2}
    \gamma 
    = \int_0^\tau\dt\,\AA\ket{\dot\psi_t}
    =\tau\langle H-\epsilon\rangle\;\mod\,2\pi.
\end{equation}
If $\epsilon$ is smaller than $\langle H \rangle$, then $\tau\langle H-\epsilon\rangle \geq \gamma$, and if $\epsilon$ is larger than $\langle H \rangle$, then $\tau\langle \epsilon - H \rangle\geq 2\pi-\gamma$. The estimate \eqref{MLQSL} follows from the observation that $\langle H-\epsilon\rangle$ assumes its smallest positive value for $\epsilon=\overline{\epsilon}$ and its largest negative value for $\epsilon=\underline{\epsilon}$.

It remains to show that $\ket{\psi_t}$ is a closed lift of $\rho_t$. Since $\epsilon$ is occupied by $\rho$ there is an eigenvector $\ket{\epsilon}$ of $H$ with eigenvalue $\epsilon$ such that $\braket{\epsilon}{\psi}\ne 0$. By adjusting the phase of $\ket{\epsilon}$, we can assume that $\ket{\psi}$ is in phase with $\ket{\epsilon}$, that is, $\braket{\epsilon}{\psi}> 0$. The curve $\ket{\psi_t}$ is a lift of $\rho_t$ since
\begin{equation}
    \ketbra{\psi_t}{\psi_t}
    = e^{-it(H-\epsilon)}\ketbra{\psi_0}{\psi_0}e^{it(H-\epsilon)}
    = e^{-itH}\ketbra{\psi}{\psi}e^{itH}
    =\rho_t.
\end{equation}
Moreover, $\ket{\psi_t}$ is in phase with $\ket{\epsilon}$ for every $t$:
\begin{equation}
    \braket{\epsilon}{\psi_t}
    = \bra{\epsilon} e^{-it(H-\epsilon)}\ket{\psi_0}
    = \braket{\epsilon}{\psi}
    >0.
\end{equation}
The initial and final vectors $\ket{\psi_0}$ and $\ket{\psi_\tau}$ sit in the fiber over $\rho$, and both vectors are in phase with $\ket{\epsilon}$. Since there is only one vector in the fiber over $\rho$ that is in phase with $\ket{\epsilon}$, the curve $\ket{\psi_t}$ is closed.

\subsection{Derivation of the Mandelstam-Tamm type bound}\label{Derivation of MTQSL}
In this section, we determine the shortest Fubini-Study length a closed curve in $\PP$ can have, given that its geometric phase is $\gamma$. To be precise, we show that this length, denoted $L(\gamma)$, equals $\sqrt{\gamma(2\pi-\gamma)}$. Equation \eqref{tjunio}, and thus the Mandelstam-Tamm type bound \eqref{MTQSL}, follows from the fact that the length of any closed curve in $\PP$ with geometric phase $\gamma$ is greater than $L(\gamma)$. If $\gamma=0$, then $L(\gamma)=0$. We therefore assume that $\gamma>0$. The calculation of $L(\gamma)$ is inspired by reference \cite{TaNaHa2005}.

We equip the total space $\SS$ of the Hopf bundle $\eta$ with the metric $g$ induced from the real part of the Hermitian inner product on $\HH$,
\begin{equation}
    g(\ket{\dot\psi_1},\ket{\dot\psi_2})
    =\frac{1}{2}\big(\braket{\dot\psi_1}{\dot\psi_2}+\braket{\dot\psi_2}{\dot\psi_1}\big). 
\end{equation}
Then $\eta$ is a Riemannian submersion, that is, $\d\eta$ preserves the inner product between horizontal vectors. As a consequence, $\rho_t$ and all of its horizontal lifts have the same length: If $\ket{\psi_t}$ is a horizontal lift of $\rho_t$,
\begin{equation}
  \length(\rho_t)
    =\int_0^\tau\!\dt\sqrt{\gfs(\dot\rho_t,\dot\rho_t)}
    =\int_0^\tau\!\dt \sqrt{g(\ket{\dot\psi_t},\ket{\dot\psi_t})}
    =\length(\ket{\psi_t}).    
\end{equation}

Let $\rho_t$, with $0\leq t\leq \tau$, be a closed curve at $\rho$ with geometric phase $\gamma$ and length $L(\gamma)$. Since geometric phase and length are parameterization invariant quantities, we may assume that $\rho_t$ has a constant speed and that $\tau=1$. Let $\ket{\psi}$ be any vector in the fiber above $\rho$, and let $\ket{\psi_t}$ be the horizontal lift of $\rho_t$ that starts at $\ket{\psi}$. Since $\rho_t$ is a closed curve with geometric phase $\gamma$, the curve $\ket{\psi_t}$ ends at $e^{i\gamma}\ket{\psi}$. Also, since $\rho_t$ has minimal length among such curves, $\ket{\psi_t}$ is a minimum for the augmented kinetic energy functional
\begin{equation}
    \mathcal{E}\big(\ket{\phi_t},\lambda_t\big)=\frac{1}{2}\int_0^1 \dt\,\big( \braket{\dot\phi_t}{\dot\phi_t}+2i\lambda_t\braket{\phi_t}{\dot\phi_t}\big),
\end{equation}
the $\lambda_t$ being a Lagrange multiplier enforcing horizontally. The kinetic energy functional is defined on curves in $\SS$ that extend between $\ket{\psi}$ and $e^{i\gamma}\ket{\psi}$.

Every variation vector field along $\ket{\psi_t}$ has the form $-iX_t\ket{\psi_t}$ for some curve of Hermitian operators $X_t$. 
To keep the endpoints of $\ket{\psi_t}$ fixed, we require that $X_t$ vanishes at $t=0$ and $t=1$. A variation with this variation vector field is $ \ket{\psi_{s,t}}=U_{s,t}\ket{\psi_t}$, with $U_{s,t}=\vec{\mathcal{T}}e^{-isX_t}$. The time order is such that $\partial_t U_{s,t} = -isU_{s,t}\dot X_t$. A straightforward calculation shows that
\begin{equation}\label{partialintegration}
	\frac{\d}{\d s}\mathcal{E}\big(\ket{\psi_{s,t}},\lambda_t\big)\Big|_{s=0} 
   	= \frac{1}{2} \int_0^1\dt \tr\Big( X_t\frac{\d}{\d t}\big(i\ketbra{\psi_t}{\dot\psi_t}- 
   	i\ketbra{\dot\psi_t}{\psi_t}  -2\lambda_t\ketbra{\psi_t}{\psi_t}\big)\Big). 
\end{equation}
Since $\ket{\psi_t}$ is a minimum of $\mathcal{E}$, and equation \eqref{partialintegration} holds for all curves of Hermitian operators $X_t$ vanishing for $t=0$ and $t=1$, the Hermitian operator
\begin{equation}\label{bee}
   A=i\ket{\dot\psi_t}\bra{\psi_t}-i\ket{\psi_t}\bra{\dot\psi_t} +2\lambda_t\ket{\psi_t}\bra{\psi_t}
\end{equation}
is time-independent. Furthermore, since $\ket{\psi_t}$ is horizontal, $i\ket{\dot\psi_t} = (A-2\lambda_t) \ket{\psi_t}$ and $2\lambda_t = \bra{\psi_t}A\ket{\psi_t}$. 
These equations imply that $\lambda_t$ is constant:
\begin{equation}
    2\dot\lambda_t
    = i \bra{\psi_t} (A-2\lambda_t) A \ket{\psi_t}-i\bra{\psi_t}A(A-2\lambda_t)\ket{\psi_t}=0.
\end{equation}
We write $\lambda_t=\lambda$ and put $B=A-2\lambda$. Then $\ket{\dot\psi_t} = -iB\ket{\psi_t}$ and $\langle B\rangle=\bra{\psi_0}B\ket{\psi_0}=0$. We show that $\rho_t=\ketbra{\psi_t}{\psi_t}$ occupies at most two eigenvalues of $B$, and we determine $\braket{\dot\psi_0}{\dot\psi_0}$, which is the speed of $\rho_t$ squared and thus equals $L(\gamma)^2$.

Since $\ket{\psi_t}$ is horizontal, the vectors $\ket{\psi_0}$ and $\ket{\dot\psi_0}$ are perpendicular. Also, by \eqref{bee}, they span the support of $A$. Any vector which is perpendicular to $\ket{\psi_0}$ and $\ket{\dot\psi_0}$ belongs to the kernel of $A$ and therefore is an eigenvector of $B$ with eigenvalue $-2\lambda$. We conclude that $\ket{\psi_0}$, and thus that the entire curve $\ket{\psi_t}$ is contained in the span of two eigenvectors of $B$. The corresponding eigenvalues are
\begin{equation}\label{eigenvalues}
    \epsilon_{\pm}=-\lambda\pm\sqrt{\lambda^2+\braket{\dot\psi_0}{\dot\psi_0}}.
\end{equation}
The requirement $\ket{\psi_1}=e^{i\gamma}\ket{\psi_0}$ is satisfied if and only if $\epsilon_+=2\pi k-\gamma$ and $\epsilon_-=2\pi l-\gamma$ for some integers $k$ and $l$. Notice that $k\geq 1$ and $l\leq 0$ since $\epsilon_+$ is positive and $\epsilon_-$ is negative. We have $-2\lambda=\epsilon_++\epsilon_-=2((k+l)\pi-\gamma$ implying that $\lambda=\gamma-(k+l)\pi$. Equation \eqref{eigenvalues} yields
\begin{equation}
    2k\pi-\gamma=-\gamma+(k+l)\pi+\sqrt{(\gamma-(k+l)\pi)^2+\braket{\dot\psi_0}{\dot\psi_0}}.
\end{equation}
It follows that $\braket{\dot\psi_0}{\dot\psi_0} = (2k\pi-\gamma)(\gamma-2l\pi)$. The right side is minimal for $k=1$ and $l=0$. We conclude that $\braket{\dot\psi_0}{\dot\psi_0}=\gamma(2\pi-\gamma)$ and hence that $L(\gamma)=\sqrt{\gamma(2\pi-\gamma)}$.

\section{Summary and outlook}\label{sec: summary}
The Aharonov-Anandan geometric phase factor is the simplest example of an Abelian quantum mechanical holonomy. We have derived three tight lower bounds on the time it takes to generate an Aharonov-Anandan geometric phase. The limits originate in geometric descriptions of the classical Mandelstam-Tamm and Margolus-Levitin quantum speed limits. In holonomic quantum computation, quantum gates and circuits are generated from more complex Abelian and non-Abelian holonomies \cite{ZaRa1999, SjToAnHeJoSi2012, AlSj2022, ZhKyFiKwSjTo2021, An1988}. In a forthcoming paper we will report lower bounds for the time it takes to generate such holonomies.

\end{document}